\documentstyle[11pt,aaspp4]{article}

\begin {document}
\slugcomment{To appear to December 1998 AJ}
\title{ The distance to NGC 6397 by M-subdwarf main-sequence fitting}

\author {I. Neill Reid}
\affil {Palomar Observatory, 105-24, California Institute of Technology, Pasadena, CA 91125,
e-mail:  inr@astro.caltech.edu}

\author {John E. Gizis}
\affil {Department of Physics and Astronomy, LGRT 532A, University of Massachusetts, Amherst,
MA 01003-4525, e-mail:  gizis@stratford.phast.umass.edu}

\begin {abstract}

Recent years have seen a substantial improvement both in photometry
of low luminosity stars in globular clusters and in modelling the
stellar atmospheres of late-type dwarfs. We build on these observational and theoretical
advances in undertaking the first determination of the distance to
a globular cluster by main-sequence fitting using stars on the
lower main sequence. The calibrating stars are extreme M subdwarfs, as
classified by Gizis (1997), with parallaxes measured to a precision of better
than 10\%. Matching against King et al's (1998) deep (V, (V-I)) photometry
of NGC 6397, and adopting E$_{B-V}=0.18$ mag, we derive a true distance
modulus of 12.13$\pm0.15$ mag for the cluster. This compares with (m-M)$_0=12.24\pm0.1$
derived through conventional main-sequence fitting in the (V, (B-V)) plane. 
Allowing for intrinsic differences due to chemical composition, we derive a 
relative distance modulus of $\delta{\rm (m-M)}_0=2.58$ mag between NGC 6397 and the
fiducial metal-poor cluster M92. We extend this calibration to other
metal-poor clusters, and examine the resulting RR Lyrae (M$_V$, [Fe/H]) relation.

\end {abstract}

\section {Introduction }

The completion and publication of the results obtained by the Hipparcos
astrometric satellite (ESA, 1997) have led to a renaissance in classical and
neo-classical astronomy. One of the latter disciplines is main-sequence fitting: estimating
the distance to globular clusters based on matching their colour-magnitude
diagrams against a fiducial sequence defined by nearby subdwarfs of similar abundance and
accurately-known parallaxes (Sandage, 1970). Several studies have applied Hipparcos
data to this subject, with most (Reid, 1997, 1998; Gratton et al, 1997; Chaboyer et al,
1998) concluding that the analysis favours larger distances, particularly for the
extreme metal-poor systems, although Pont et al (1998) recover the
pre-Hipparcos results for M92. Combined with recent revisions in stellar
models, however, even the last study leads to ages of less than 14 Gyrs for the oldest clusters,
while the longer distance-scale results point to ages of 11 to 13 Gyrs.

Despite the addition of Hipparcos data, there are relatively
few upper main-sequence (FG early-K) subdwarfs with parallaxes determined to
an accuracy of better than 10\%. This reflects the scarcity of halo stars
near the Sun: with a space density of $3 \times 10^{-6} \ {\rm stars pc}^{-1} \
{\rm M}_V^{-1}$,
one expects only 40-50 such stars within 100
parsecs. Not all of those subdwarfs have Hipparcos data and many which 
were observed are sufficiently faint that parallaxes are measured 
to an accuracy of only 1.5 to 2 milliarcseconds. In particular, the Hipparcos
dataset includes very few extreme  subdwarfs ([Fe/H]$<$-1.5) with both
reliable parallaxes and reliable abundance determinations. 

By analogy with the disk, one expects the number density of halo subdwarfs to
increase with decreasing mass, and surveys of both globular clusters (Fahlman et al, 
1989; Paresce, Demarchi \& Romaniello, 1995) and the field (Dahn et al, 1995) 
show that this is indeed 
the case. Follow-up observations of proper-motion stars, particularly those drawn from
the Luyten Half-Second catalogue (LHS - Luyten, 1980) have resulted in the
identification of several dozen late-K and M-type subdwarfs within 50-100 parsecs of
the Sun. Most of those stars have sufficiently faint apparent magnitudes to
permit ground-based CCD parallax measurements, which can achieve sub-milliarsecond
precision (Monet et al, 1992). Moreover, with such faint
{\sl apparent} magnitudes, the transformation from relative to absolute 
parallaxes is robust. 

In principle, M-subdwarfs could also be used as templates
for main-sequence fitting distance determination. Until recently, there have 
been two substantial obstacles: first, globular cluster colour-magnitude
diagrams were defined poorly at M$_V > 8$; second, reliable abundance estimates 
for late-type subdwarfs lay beyond the scope of stellar atmosphere models.
The advent of WFPC2 on the Hubble Space Telescope has eliminated the
first obstacle, while abundances in M-dwarfs can be at least constrained 
using the extensive theoretical analyses by Allard \& Hauschildt (1995).

Taking advantage of these advances, this paper presents the first attempt to use 
the main-sequence defined by nearby extreme M-subdwarfs to estimate the
distance of a metal-poor globular cluster. Section 2 outlines our
calibration and selection of the appropriate local reference stars, and section
3 matches those stars against the deep HST observations of NGC 6397 obtained
by King, Anderson, Cool \& Piotto (1998:KACP). The final section summarises the future prospects
for this technique.

\section {The subdwarf calibrators}

The existence of subluminous stars, lying between the disk main sequence and
white dwarfs, was first suggested by Adams \& Joy (1922), who
identified three weak-lined "A-type" stars (HD 19445, HD 219617 and HD 140283) with
unusual absolute magnitudes, while the
actual term subdwarf was coined by Kuiper (1939). Sandage \& Eggen (1959), however, were
the first to demonstrate that these stars, subluminous in the observational 
(M$_V$, (B-V)) and (M$_V$, spectral type) planes, also fell below the main sequence 
in (M$_{bol}$, T$_{eff}$). Most early examples were drawn from proper motion
surveys. Since the Galactic halo is a high velocity dispersion, low rotation
population, the typical subdwarf has a velocity of 200-250 kms$^{-1}$ with
respect to the Sun. As a result, while the local disk to halo number ratio
is $\approx 400:1$ by volume, the subdwarf contribution to catalogues of
high proper-motion stars can exceed 25\%.

Initial studies concentrated on the more luminous F and G-type stars, although
later-type high velocity stars have been known for over 100 years - 
Kapteyn's star (Gl 191 - M$_V$=8.6) was identified in 1897.
Later-type (K and M) subdwarfs are characterised by strong metal hydride absorption,
notably MgH (Greenstein, 1971; Cottrell, 1978) and CaH (Jones, 1973; Mould, 1976).
Over thirty such stars now have accurate trigonometric parallax measurements, 
providing an empirical
description of the lower main-sequence in the HR diagram of metal-poor stars (Gizis, 1997).

As one decreases the metal abundance of an M-type dwarf, TiO bands decrease
in strength relative to metal hydride bands. This reflects partly the double-metal
nature of TiO, partly the fact that titanium is in competition with hydrogen (via
H$_2$O) for a decreasing supply of oxygen atoms (Mould, 1976). Reid, Hawley \& Gizis (1995)
devised a series of narrowband indices to characterise the behaviour of both
TiO and CaH, allowing subdwarfs to be ranked in terms of relative abundance. Gizis (1997)
has used those indices to classify subdwarfs into two categories: intermediate (sdM)
and extreme (esdM) subdwarfs (figure 1). One star stands out amongst the latter - LHS 453,
with CaH1$\sim0.51$ but effectively no TiO absorption. The optical spectrum indicates
that the abundance  is substantially
lower than that of the average esdM (Gizis, 1997), but the position on the colour-magnitude
diagram (M$_V$=13.08, (V-I)=2.22) is not anomalous.

Transforming relative bandstrength measurements to metallicities requires
appropriate model atmospheres. Gizis matches spectra spanning the wavelength range
$\lambda\lambda 6200-7400 \AA$, including the CaH bands at $\lambda 6350$ and $\lambda 6800$,
against models from the Extended grid calculated by Allard \& Hauschildt (1995).
Based on that comparison, sdM subdwarfs appear to have abundances in the range
-1 to -1.5 dex, while esdMs fall in the range -1.5 to -2.5  dex, with an
average abundance close to -2.0 dex. The
latter stars should therefore be the local analogues of the M subdwarfs found in
the extreme halo clusters, such as NGC 6397, M15 and M92.

Support for this calibration comes from spectroscopy of low-luminosity common proper motion
companions of three early-type subdwarfs of known abundance (Gizis \& Reid, 1997b). In addition, 
one can compare the relative numbers of sdM and esdM stars against the [M/H]
distribution derived by Laird et al (1987) for F and G field subdwarfs. The latter
distribution peaks at [M/H]$\sim-1.5$, with a substantial tail to lower abundance, and one
expects a similar distribution amongst lower-luminosity stars. Clearly, the currently-available
sample of late-type subdwarfs with both accurate parallax measurements and spectroscopic 
observations is neither complete nor volume-limited, but since most attention has been 
devoted to stars lying well below the main-sequence, one expects any bias to favour
selection of the most metal-poor systems. 

Following those arguments, we have identified the stars listed in Table 1 as local
calibrators, defining a reference lower main-sequence for the metal-poor ([M/H]$<-1.5$)
halo. All are spectroscopically confirmed esdM stars (Gizis \& Reid, 1997a; Reid \& Gizis, in prep.).
Figure 2 plots the (M$_V$, (V-I)) colour-magnitude diagram described by
these stars, together with data for disk dwarfs drawn from the 8-parsec sample 
- the volume-limited sample of all stars north of $\delta = -30^o$ which
are currently known to be within 8-parsecs of the Sun (Reid \& Gizis, 1997).

Lutz-Kelker corrections (Lutz \& Kelker, 1973) are usually invoked when combining 
trigonometric parallax data to derive statistical relations, such as determining a 
mean colour-magnitude relation or in main-sequence fitting. The scale of those
corrections depends both on the precision of the individual measurements, and on
the distribution, N($\pi$), of the parent population. The latter can
be regarded as the true density distribution as modified by parallax-independent
selection effects. Table 1
shows that most of the stars in our reference sample have parallaxes measured to a
precision of better than 10\%, corresponding to systematic corrections of $< 0.1 $mag.
for a uniform space distribution. The latter corrections are listed as $\Delta LK$ in Table 1. 
However, since all of the calibrating subdwarfs were identified 
in surveys of high proper-motion stars, the underlying distribution N($\pi$) is
far from uniform. One expects an initial rise as $\pi^{-4}$, but a subsequent flattening 
and turnover at $\pi_{crit}$, whose value depends on the proper motion limit and
the subdwarf velocity distribution, and appropriate LK corrections will therefore be
smaller than for N($\pi) \propto \pi^{-4}$. To be conservative, we 
apply no Lutz-Kelker corrections in our main-sequence fitting analysis.

\section {The distance to NGC 6397}

NGC 6397 provides the best target for distance determination through main-sequence
fitting using late-type subdwarfs. The cluster is one of the closest to the
Sun, lying at Galactic co-ordinates (l=338$^o$.2, b=-12$^o$.0), having an abundance of [M/H]=-1.82.
(Carretta \& Gratton, 1997). The latter metallicity estimate is based on 
spectroscopy of ten red giant stars,
and is 0.34 dex higher than the same authors' calibration of M92, the fiducial
extreme metal-poor cluster (Bolte \& Hogan, 1995). Most importantly, NGC 6397 has
deep imaging data obtained by the Hubble Space Telescope, extending to close
to the hydrogen-burning limit (Cool, Piotto \& King, 1996). The low latitude
of NGC 6397 leads to substantial contamination of the colour-magnitude diagram
by field stars. However, the HST observations were made over a period of 3 years, and
that relatively short baseline, coupled with the high spatial resolution of
the HST images and the significant transverse motion of NGC 6397, has allowed KACP
to identify a proper-motion selected sample of cluster members,
providing a colour-magnitude diagram of unparalleled accuracy.

KACP's observations were made using the WFPC2 F555W and F814W filters,
whereas our calibrating subdwarf observations are on the Cousins VI system.
Fortunately, the two photometric systems have similar spectral response. 
Holtzman et al (1995) provide analytic transformations between the natural
WFPC system and the UBVRI passbands, but those are limited to (V-I)$<$1.5. 
However, they also compute synthetic transformations by convolving the 
filter transmission curves with spectrophotometry of dwarfs, drawn from the
Bruzual et al spectral atlas, with (V-I) colours as red 3.0 magnitudes.
As yet, there are no empirical tests of those predicted relations for M
subdwarfs, but we are currently undertaking an HST snapshot programme, 
searching for low-luminosity companions to such stars. Our observations
are made using the F555W and F850LP filters, so we can test directly the
transformations for the former, and indirectly assess the accuracy of
the Holtzman et al curves for I-band data.

To date, we have observations of six subdwarfs: LHS 169, 174, 216, 
407, 522 and 320. All have ground-based V,I data (Gizis, 1997), with
(V-I) colours accurate to $\sim0.05$ magnitudes. Comparing magnitudes 
derived from the HST F555W data, calibrated using STSDAS, against the
ground-based data we derive $\delta V = F555W - V = 0.057 \pm 0.03$ magnitudes,
where the uncertainty quoted is the standard deviation of residuals. 
Holtzman et al predict a strong colour term between the F850LP magnitudes
and the standard Cousins I-band. Applying their specified calibration to
derive I$_{850}$, we find $\delta I = I_{850} - I = 0.012 \pm 0.06$. Given that
some of the ground-based I-band data are transformed from I$_K$, this larger
scatter is not unreasonable. These comparisons therefore suggest that M
subdwarfs do not exhibit pathological colour terms in the transformation
between the HST and ground-based broad-band photometric systems. 
In the range 1.7 $<$(V-I)$<$2.3 spanned by the M92 stars, the colour terms
predicted by Holtzman et al 
are almost parallel in (V-F555W) and (I-F814W). We have therefore take the
instrumental (F555W-I814W) colours as our best estimate of (V-I), and adopt
V=F555W-0.06 magnitudes.

Accurate accounting of line-of-sight reddening is implicit in the derivation of
the distance modulus, apparent or true, of a globular cluster by main-sequence
fitting. Since NGC 6397 lies at low latitude, it is subject to significant 
foreground reddening, although previous estimates of this quantity are internally consistent.
Most are based on the observed properties of blue horizontal branch
stars: Graham \& Doremus (1968) set a lower limit of E$_{B-V} \ge 0.16$; Newell et al
(1969) derived E$_{B-V} = 0.20\pm0.02$; Cannon (1974) derives E$_{B-V}=0.18\pm0.02$,
based on a direct comparison with BHB stars in NGC 6752; van den Bergh (1988) finds
E$_{B-V}=0.19\pm0.02$; and Heber \& Kudritzki (1986) require E$_{B-V}=0.20$ in
matching the ultraviolet spectrum of the sdO stars, ROB 162. 

Other techniques of reddening estimation 
give similar results: Alcaino et al (1996) estimated E$_{B-V}=0.17$ mag., based on 
matching the (U-B)/(B-V) distribution against data for NGC 1841 
(E$_{B-V} = 0.17$); Cannon (1974) derives E$_{B-V}=0.18$ using the same technique, but
calibrated against the population I two-colour relation; and Vandenberg, Bolte \& Stetson
match the main-sequence turnoff against M92 data to derive a differential reddening of
+0.17 magnitudes, or E$_{B-V}=0.19$. Note that the last calculation does not take into
account possible intrinsic colour differences between the two clusters, a matter we
return to in the following section. Finally, Schlegel et al. (1998) have used IRAS and DIRBE data
to derive extinction maps. Their estimated reddening in the direction of NGC 6397
is E(B-V) = $0.185 \pm 0.030$, in excellent agreement with other estimates.  

Considering all of these results, we adopt a value of E$_{B-V}=0.18$ for 
the foreground reddening toward NGC 6397. Vandenberg et al (1990) find evidence
for some variation across the face of the cluster, but at a level of $<0.015$mag.
There is no evidence that the ratio of total to selective extinction, R, is 
enhanced (as is the case with M4), so we adopt A$_V$=0.56 mag, and a ratio
${{{\rm E}_{V-I}}\over {{\rm E}_{B-V}}} = 1.35$ (Drukier et al, 1993; He et al, 1995),
giving E$_{V-I}=0.24$. The de-reddened colour-magnitude diagram for the cluster
is shown in figure 3.

Pre-Hipparcos main-sequence fitting analyses indicated a true distance
modulus for NGC 6397 of (m-M)$_0$=11.5 to 11.7 magnitudes (Anthony-Twarog, 
Twarog \& Suntzeff, 1992), but those estimates were tied primarily to one 
reference star, HD 64090. The Hipparcos parallax for that star, $\pi = 35.29 \pm 1.04$ mas, 
is 20\% smaller than the previously-accepted ground-based measurement, leading to 
an increase of at least 0.4 magnitudes in the inferred cluster distance modulus. Based on
eight subdwarfs with $ -1.55 > [M/H] > -2.1$ and with Hipparcos parallax measurements with a precision
${\sigma_\pi \over \pi} < 12 \%$, Reid (1998) derives (m-M)$_0 = 12.24\pm0.1$.

We have limited our current analysis to NGC 6397 stars in the color 
range $1.75 \le V-I \le 2.50$. This encompasses the majority of
field esdMs, but excludes both  
K subdwarfs, for which accurate metallicities are not available,
and the few very low-luminosity esdMs, such as  LHS 1742a.
Even with the KACP proper motion data, some non-members appear in the
cluster color-magnitude diagram (Figure 3).  We have identified
the main-sequence locus by a "robust" linear fit to the KACP data
with $V-I>1.5$.  This minimises the absolute value of
the deviation of each point, rather than the square of deviation,
and therefore gives little weight to 
outliers (Press et al. 1992).  All objects within $\pm 0.4$ magnitudes
of the best-fit line are accepted.  As can be seen in figure 3, 
the main sequence locus is recovered.

We then use least-squares fitting to describe the NGC6397 main
sequence in the range $1.75 \le V-I \le 2.50$.  The best-fit third-order
polynomial is

\begin{displaymath}
V \qquad = \qquad 24.517 + 3.436 ((V-I) - 2) -1.207 ((V-I) - 2)^2
\end{displaymath}

As reference stars, we select the local esdM with parallaxes of precision
better than 10\%, corresponding to uniform-density Lutz-Kelker corrections of
less than 0.1 magnitudes. If we fit a linear relation to the ten esdM in Table 1
which meet this criterion, we derive
\begin{displaymath}
M_V \qquad = \qquad 12.429 \pm 0.076 + (2.909 \pm 0.338) \times ((V-I) -2)
\end{displaymath}
which has a slope consistent with the NGC6397 data and implies a distance 
modulus of (m-M)$_0$=12.04 at $V-I=2.0$. Alternatively, we can adopt the shape of the
main-sequence defined by the NGC 6397 stars, and fit to the parallax stars, 
allowing only the zeropoint (evaluated at $V-I=2.0$) to vary.  The 
resulting zeropoint is 12.403, implying a true distance modulus of 12.13 magnitudes. 
We have also fitted the cluster colour-magnitude diagram with second- and
fourth-order polynomials, and using the same technique, derive distance moduli of
12.12 and 12.14 respectively. 

The main uncertainties in this distance determination centre on the relative
abundance of NGC 6397 and the {\sl mean} of the esdM calibrators, and on how well-matched
are the VI photometric systems. In the latter case, a systematic offset of 
$\delta$(V-I)=$\pm0.03$ mag in the colour transformation
corresponds to $\pm0.1$ mag in $\delta$(m-M); in the former case,
the metal-poor stellar models computed by Baraffe et al (1998) indicate an
offset of $\sim 0.3$ magnitudes in M$_V$ between [m/H]=-1.5 and -2.0 at
a colour of (V-I)=2.1 magnitudes. Thus a systematic error  of $\pm0.25$ dex in
the mean abundance likely corresponds to $\pm0.15$ magnitudes in distance
modulus. 

Our estimate of the
distance modulus of NGC 6397, derived by subdwarf-fitting to the lower main-sequence, is
(m-M)$_0$=12.13$\pm0.15$ mag. Averaging this result with the (M$_V$, (B-V))
analysis, with double weight assigned to the latter result, gives (m-M)$_0=12.20$.
The corresponding absolute magnitude at the turnoff is M$_V$=3.8$\pm0.1$, implying 
an age of $\sim 11$ Gyrs if matched against either the models computed by D'Antona et al
(1997) or by Bergbusch \& Vandenberg (1992).

\section {The distance to M92}

M92 has been adopted generally as the fiducial cluster for the group of
extreme halo globulars, including M15, M30 and M68, largely on account of 
its low foreground reddening. The standard estimate of E$_{B-V}=0.02$ mag.
is confirmed by Schlegel et al. (1998), who derive $0.022 \pm 0.003$. 
The available HST data for M92 do not
extend to sufficiently faint magnitudes to allow a direct distance estimate
using the techniques outlined above. However, we can use the methods outlined
by Vandenberg et al (1990) to derive a relative distance between M92 and NGC 6397.

Vandenberg et al's study was aimed primarily at probing the relative ages of
globular clusters, using the observed distribution in the colour difference
between the turnoff and the base of the giant branch in clusters of similar
abundance. Aligning the upper main sequence and subgiant branch in clusters
also gives a direct estimate of the difference in reddening, from $\Delta$B-V,
and in apparent distance modulus, from $\Delta$V, provided one assumes that 
the intrinsic colour and absolute magnitude at the turnoff are identical. 
Vandenberg et al adopt the latter assumption in comparing M92 and NGC 6397,
but that is not likely to be the case if the clusters have difference abundances.
However, we can use theoretical stellar evolutionary tracks to account for the offset.

Carretta \& Gratton's (1997) red giant analyses indicate a metallicity difference
of $\delta$[Fe/H] = 0.34 dex (all differences are expressed in the sense (M92)-(NGC6397)).
King, Stephens, Boesgaard \& Deliyanis (1998:KSBD) derive a much lower abundance 
estimate for M92 (and, from Na I absorption, a reddening of E$_{B-V}=0.05-0.07$) based
on high-resolution spectroscopy of subgiants. The high reddening is at odds 
with other studies, and there is no comparable abundance analysis of NGC 6397. 
Since we are concerned here with the relative metallicity of the two clusters,
we adopt the Carretta \& Gratton result. Vandenberg et al's figure 3 shows that
their colour-difference age-calibration technique is relatively insensitive to
metallicity at metallicities below $\sim -1.5$ dex. Thus, the general
agreement between the M92 and NGC 6397 fiducial colour-magnitude relations
is {\sl consistent} with there being no significant difference in age
between the two clusters.

Given similar ages, we expect the higher-abundance cluster, NGC 6397, to
have a fainter, redder turnoff than M92. We can, however, use theoretical
isochrones to allow explicitly for this effect in estimating the relative
distance modulus. Defining  $\delta$(B-V) and $\delta M_V$  as the 
intrinsic offset in colour and magnitude at the turnoff induced by the
metallicity difference, the model calculations by D'Antona et al (1997)
predict values of 
\begin{displaymath}
\delta (B-V) \qquad = \qquad -0.015 \quad {\rm mag}
\end{displaymath}
\begin{displaymath}
\delta M_V \qquad = \qquad -0.07 \quad {\rm mag}
\end{displaymath}
for the -0.34 dex NGC 6397/M92 abundance difference. 

The offsets $\Delta (B-V)$ and $\delta V$,
derived through the Vandenberg et al technique of aligning cluster fiducial
diagrams at the turnoff, can now be expressed in terms of differences in
distance modulus, foreground reddening and in the intrinsic cluster
properties, as follows,
\begin{displaymath}
\Delta {\rm(B-V)} \qquad = \qquad \delta {\rm E_{B-V}} \quad + \quad \delta (B-V)
\end{displaymath}
\begin{displaymath}
\Delta V \qquad = \qquad \delta (m-M)_0 \quad - \quad R \delta {\rm E_{B-V}} \quad - \quad \delta M_V
\end{displaymath}
Given measured values of $\Delta$(B-V)=-0.17 mag and $\Delta$V=2.17 mag for
the M92/NGC 6397 pairing, adopting the 
standard value of R=3.12 leads to an estimate of 2.58 magnitudes for the difference in true
distance modulus between M92 and NGC 6397. In section 3 we derived an
averaged value of (m-M)$_0 = 12.20\pm0.1$ for NGC 6397, giving 
$\langle$(m-M)$_0\rangle = 14.78$ mag for M92, with an estimated 
uncertainty of $\pm0.1$ magnitude. Again, matching the turnoff absolute magnitude
against either the D'Antona et al or Bergbusch \& Vandenberg models leads to 
similar age estimates of 12 to 13 Gyrs. 

Comparing the current M92 distance modulus with other recent determinations, Bolte
\& Hogan's (1995) semi-empirical analysis, based on ground-based parallax data (notably
for HD 103095), gives (m-M)$_0=14.65$ mag, while D'Antona et al (1997 - echoed by
Reid, 1997) derive (m-M)$_0=14.80$ using a larger sample of reference stars. Of the
post-Hipparcos analyses, Reid (1997, 1998) derives (m-M)$_0=14.94$, but the 
calibrating subdwarfs in the former analysis include binaries and are calibrated on the
Carney et al (1994) abundance scale, while the value cited in the latter paper 
uses Vandenberg et al's NGC 6397 ($\Delta$V, $\Delta$(B-V)) method, but without correction 
for cluster to cluster metallicity differences.
Gratton et al (1997) find (m-M)$_V=14.82\pm0.08$ (for E$_{B-V}$=0.03), while 
Chaboyer et al's (1998) distance modulus, based on eight calibrating
subdwarfs, is (m-M)$_0=14.76$. Finally, Pont et al's result is (M-M)$_0=14.68$ if one
excludes all potential binaries from their reference sample. Thus, the individual
studies span a range of less than 0.15 magnitudes in true distance modulus. Moreover, 
as emphasised in the introduction, even the shortest distance estimate implies a cluster
age of only$\sim13$ Gyrs, marking a significant easing of the cosmological constraints
highlighted by Bolte \& Hogan.

\section {The RR Lyrae (M$_V$, [Fe/H]) relation}

Several recent studies have revisited the field RR Lyrae (M$_V$, [Fe/H]) calibration
(Layden et al, 1996; Fernley et al, 1998; Gould \& Popowski, 1998), and essentially re-confirmed the 
results originally derived by Strugnell et al (1986) and Barnes \& Hawley (1987):
a mean absolute magnitude of M$_V$=0.77$\pm$0.15, independent of [Fe/H]. Gould \&
Popowski and Udalski (1998) suggest that those results support a short distance scale ((m-M)=18.1 to 18.2)
to the LMC, and argue for a systematic error in globular cluster distances derived 
by main-sequence fitting. NGC 6397 has no RR Lyrae population, but the instability strip
is populated in the extreme metal-poor systems, notably M15.
We briefly re-examine this issue based on the present results.

In the previous section we derived the reddening and distance modulus of M92 
relative to NGC 6397, making due allowance for the abundance differences. 
Vandenberg et al originally applied this same technique in estimating
($\Delta$V, $\Delta$(B-V)) for a larger sample of cluster, choosing reference
clusters of the appropriate metallicity. In particular, they calibrate M68 relative
to M92 and M3 relative to NGC 6752 ((m-M)$_0$=13.16, E$_{B-V}$=0.04; Reid, 1998).
We have used the same method to estimate the relative distance modulus between M15
and M92. In addition, M5 has direct main-sequence fitting distance estimates by Gratton
et al (1997) and Reid (1998) based on Hipparcos subdwarfs, and we have averaged those 
to derive (m-M)$_0$=14.50.\footnote{As emphasised by Reid (1998), calibrating the distance
to M5 based on HD 103095 ([Fe/H]=-1.22) alone gives (m-M)$_0$=14.43.}
All of those clusters have substantial RR Lyrae populations, and Table 2 lists the
corresponding mean absolute magnitude estimates for the variable star
population. 

The five clusters listed in Table 2 span a range of 1 dex in abundance, and allow 
an estimate of the metallicity dependence of the absolute magnitudes of RR Lyraes.
Those data are plotted in figure 4, and compared with Layden et al's 
statistical parallax results; Sandage's (1982, 1993) cluster variable calibration, 
where the gradient ${dM_V \over d[Fe/H]} \sim 0.3$ is derived from period-shift
analysis; Cacciari et al's (1992) and McNamara's (1997) field-star Baade-Wesselink 
analyses; and the
relation predicted by Caloi et al's (1997) theoretical horizontal branch models
(plotted from their Table 2 for log$T_{eff}=3.83$). Cassisi et al (1998) have
recently computed OPAL-based models which are in good agreement with the latter,
favouring a slightly steeper M$_V$/[Fe/H] relation.
It is clear that the main-sequence 
fitting calibration is in close agreement with Sandage's analysis, McNamara's recent
Baade-Wesselink calibration and with Caloi et al's theoretical models.

Comparing specifically against the statistical parallax results, at intermediate 
abundances the discrepancy lies at the 1$\sigma$ level, but is more significant 
at [Fe/H]=-2. Were we to adopt Gould \& Popowski's uniform calibration, M$_V$(RR)=0.77, 
the implied distance modulus of M92 is only 14.27 magnitudes. 
This is 0.35 to 0.4 magnitudes lower than the {\sl shortest} distance estimate (Pont 
et al, 1998) based on main-sequence fitting. Given the predominance of BHB stars in M92,
evolutionary effects may lead to higher luminosities for that clusters RR Lyraes
(Lee et al, 1990), as
suggested by the main-sequence fitting results, but, with 
well-populated instability strips, such considerations are not relevant to M15 and M68.
Udalski does not
discuss this issue; prompted partly by KSBDs results for M92, Gould \& Popowski suggest 
that the origin may lie in a mismatch of the subdwarf and red giant cluster abundance scales.
However, adjusting (m-M) by 0.35 magnitudes in
main-sequence fitting corresponds to a change of -0.07 mag in (B-V) in the colours of
the reference subdwarfs. In the case of M15 or M92, that adjustment implies an
extremely low abundance. for example, the D'Antona et al models predict
${\partial{(B-V)} \over {[Fe/H]}} \sim 0.05$ mag dex$^{-1}$ at [Fe/H]=-2, and that
gradient should decrease with decreasing abundance and decreasing line-blanketting.
Hence a correction of -0.07 magnitudes in (B-V) is likely consistent with
$\delta$[De/H]$\sim -2$, and an abundance of -4 dex for M92 seems extremely unlikely. 

An alternative explanation, which accounts at least for the discrepancy at
the lowest metallicities, may lie in the abundance distribution of the field
star sample (figure 4). Only twenty-two stars in the Layden et al sample have [Fe/H]$<-2$.
On the other hand, the latter authors derive $\langle M_V \rangle = 0.73\pm0.18$
for a sample of 83 variables with $\langle [Fe/H] \rangle = -1.83$. Moreover, this
proposal does not address the 0.2 to 0.3 magnitude discrepancy at [Fe/H]$\sim-1$,
emphasised by Reid (1998), between M5 and  the Galactic Centre variable stars.
Thus, the conflict between the absolute magnitude calibrations defined by cluster
and field RR Lyraes remains to be resolved.

\section {Summary and conclusions}

We have used observations of nearby esdM subdwarfs to define the location of
the lower main-sequence of the metal-poor halo in the (M$_V$, (V-I))
colour-magnitude diagram. Matching that calibrated sequence against deep HST
photometry of NGC 6397, we estimate a cluster distance modulus of 12.12$\pm$0.15
magnitudes for a foreground reddening of E$_{B-V}=0.18$ magnitudes. This is
in excellent agreement with Reid's (1998) estimate of (m-M)$_0=12.24$, based
on traditional main-sequence fitting in the (M$_V$, (B-V)) plane using
earlier-type subdwarfs with Hipparcos parallax data, and supports the larger
distance of $\sim 2750\pm150$ parsecs. 

We have combined the predictions of theoretical evolutionary models with
Vandenberg et al's measurements of relative offsets in the (V, (B-V)) plane
to estimate a distance to M92. Making allowance for the difference in
abundance between the two clusters, we estimate M92 to lie at a distance 
modulus of 14.78$\pm$0.1 magnitudes, a relatively modest increase over 
pre-Hipparcos determinations. Matched against the recent set of 
metal-poor stellar models computed by D'Antona et al (1997), both clusters
are estimated to have ages of $\sim 12$ Gyrs. Extending the distance
calibration to M15, M68 and M3, we derive absolute magnitudes
for RR Lyrae variables which are in excellent agreement with both
Sandage's (M$_V$, [Fe/H]) relation and the predictions of the recent
horizontal branch models by Caloi et al (1997). 

The principal aim of this paper is to demonstrate that main-sequence
fitting distance estimation need no longer be confined to using parallax
stars on the upper main sequence. Deep, high spatial-resolution photometry
with the Hubble Space Telescope already provides observations reaching
the hydrogen-burning limit in the nearer globular clusters, while the
most recent stellar atmosphere calculations allow reliable abundance
estimates for the coolest subdwarfs. Further refinements in the models
combined with observations with the future Next-Generation Space Telescope
will allow the full exploitation of this latest variation on a neo-classical theme.

\acknowledgments 
We would like to thank Messrs. King, Cool, Piotto and Anderson
for making available the full photometric catalogue they have compiled of 
proper-motion members of NGC 6397. This work was supported partially by
NASA grant GO-07385.01-96A from the Space Telescope Science Institute,
which is operated by the Association of Universities for Research in
Astronomy, Incorporated, under NASA contract NAS5-26555.

\begin{deluxetable}{rcccccccccr}
\tablewidth{0pt}
\tablenum{1}
\tablecaption{esdM parallax subdwarfs}
\tablehead{
\colhead{LHS } & \colhead{name} &  \colhead{M$_V$} &\colhead{(V-I)} 
&\colhead{class} & \colhead {$\pi$(")} & \colhead {$\sigma_\pi \over \pi$}
&\colhead {$\Delta LK$} }
\startdata
169 & Gl 129 & 11.58 & 1.72 & esdK7& 0.0309 & 0.074 & -0.06  \\
192 & LP 302-31 & 12.37 & 1.98 & esdM1& 0.0102 & 0.078 & -0.06 \\
364 & G165-47 & 12.47 & 1.95 & esdM1.5&0.0374 & 0.099 & -0.10 \\
375 & LP 857-48 & 13.66 & 2.20 & esdM4& 0.0395 & 0.025 &-0.01 \\
453 & LP139-13 & 13.08 & 2.22 & esdM3.5& 0.0103 & 0.087 & -0.08 \\
522 & Gl 861 & 11.29 & 1.62 & esdK7& 0.0268 & 0.078 & -0.06 \\
1174 & LP 406 & 12.97 & 2.09 & esdM3& 0.0157 & 0.076 & -0.06 \\
1742a & LP 417-42 & 14.44 & 2.74 & esdM5.5 & 0.0134 & 0.090 & -0.08 \\
1970 & LP 484-6 & 13.31 & 2.09 & esdM2.5 & 0.0129 & 0.062 & -0.04 \\
2045 & LP 545-45 & 13.72 & 2.46 & esdM4.5 & 0.0111 & 0.081 & -0.07 \\
3061 & LP 502-32 & 14.25 & 2.61 & esdM5& 0.0089 & 0.090 & -0.08 \\
3382 & LP 24-219 & 12.11 & 2.09 & esdM2.5 & 0.0104 & 0.087 & -0.08 \\
3390 & LP 181-51 & 13.87 & 2.48 & esdM4.5&0.0123& 0.081 & -0.07  \\
3548 & LP 695-96 & 12.12 & 2.09 & esdM3 & 0.0083&  0.072 & -0.05 \\
3628 & LP 757-13 & 12.12 & 2.05 &esdM1.5& 0.0088 & 0.090 & -0.09 \\
\enddata
\tablecomments{ Parallax data from Monet et al (1992) except \newline
LHS 169, 364, 522 -- van Altena et al, 1996 \newline
LHS 375 -- Ruiz \& Anguita, 1993}
\end{deluxetable}

\begin{deluxetable}{lrrcccrrrrrrrrrr}
\footnotesize
\tablecaption{RR Lyrae absolute magnitudes}
\tablewidth{0pt}
\tablenum{2}
\tablehead{
\colhead{cluster } & \colhead{ E$_{B-V}$ }& \colhead{[Fe/H]} &
\colhead {$\Delta$V} & \colhead {$\Delta$(B-V)} & \colhead {reference} &
\colhead {(m-M)$_0$} & \colhead{ $\langle V_0 \rangle$(RR}) &
\colhead { M$_V$ (RR)} & \colhead{RR Lyrae data} }
\startdata
M5 & 0.02 & -1.11 & 0 & 0 & M5 & 14.50 & 15.00 & 0.50 & Reid (1996) \nl
M3 & 0.03 & -1.34 & -1.75 & 0.01 & NGC6752 & 14.88 & 15.54 & 0.66 & Buonnano et al (1994) \nl
M68 & 0.05 & -1.99 & -0.43 & -0.033 & M92 & 15.11 & 15.43 & 0.32 & Walker, 1994 \nl
M15 & 0.09 & -2.12 & -0.57 & -0.07 & M92 & 15.23 & 15.55 & 0.32 & Silbermann \& Smith, 1995 \nl
M92 & 0.02 & -2.16 & 0 & 0 & M92 & 14.78 & 15.04 & 0.26 & Carney et al, 1994 \nl
\enddata
\tablecomments{Matching M3 against the upper main-sequence of M13 leads to
an inferred distance modulus of 15.10 magnitudes}
\end{deluxetable}

\begin{figure}
\plotfiddle{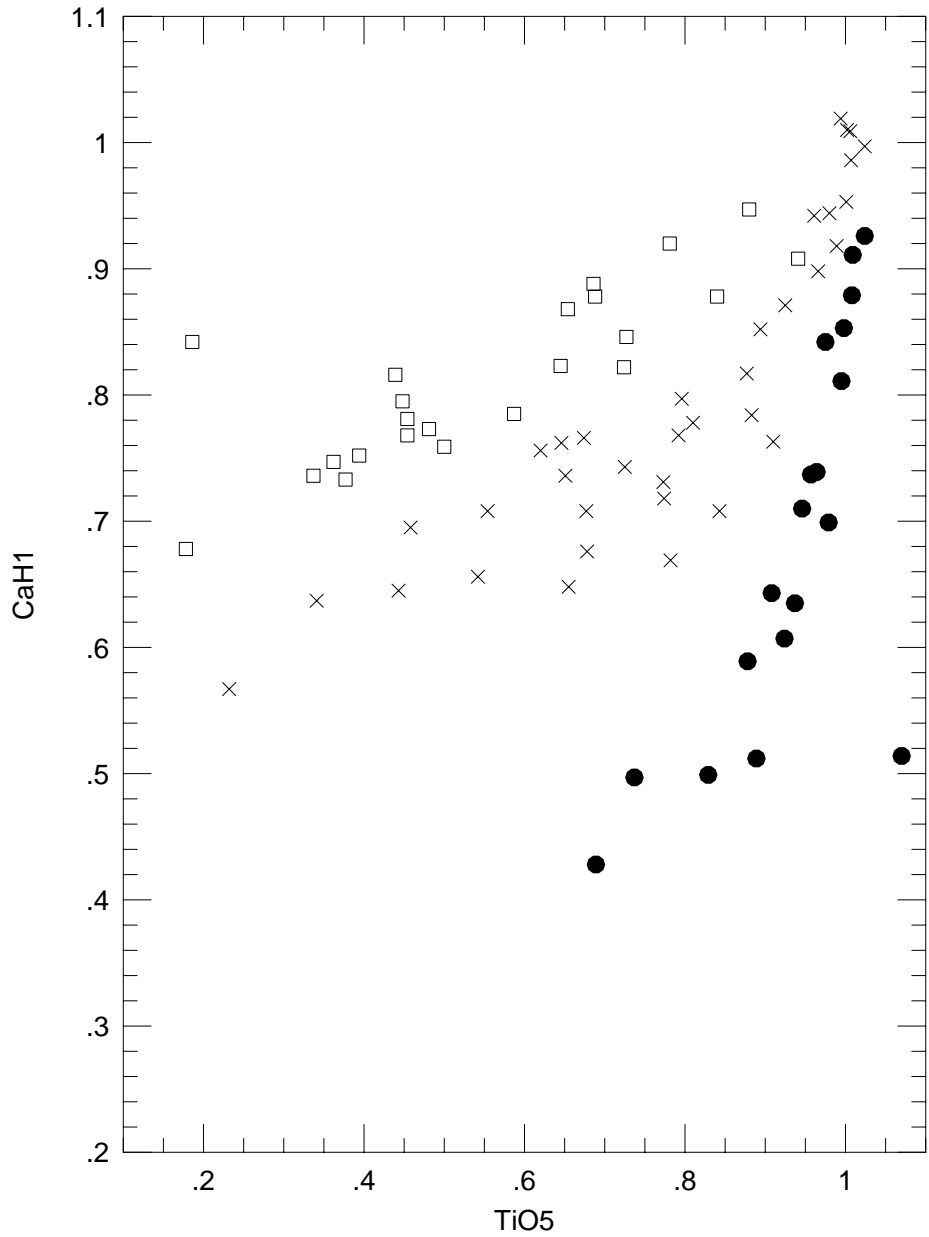}{6in}{0}{70}{70}{-230}{-30}
\caption{Identifying M subdwarfs - the calibration of solar-abundance disk dwarfs (open
squares), intermediate-abundance sdM subdwarfs (crosses), and extreme subdwarfs, esdM, 
in terms of the relative strength of CaH and TiO absorption}
\end{figure}

\begin{figure}
\plotfiddle{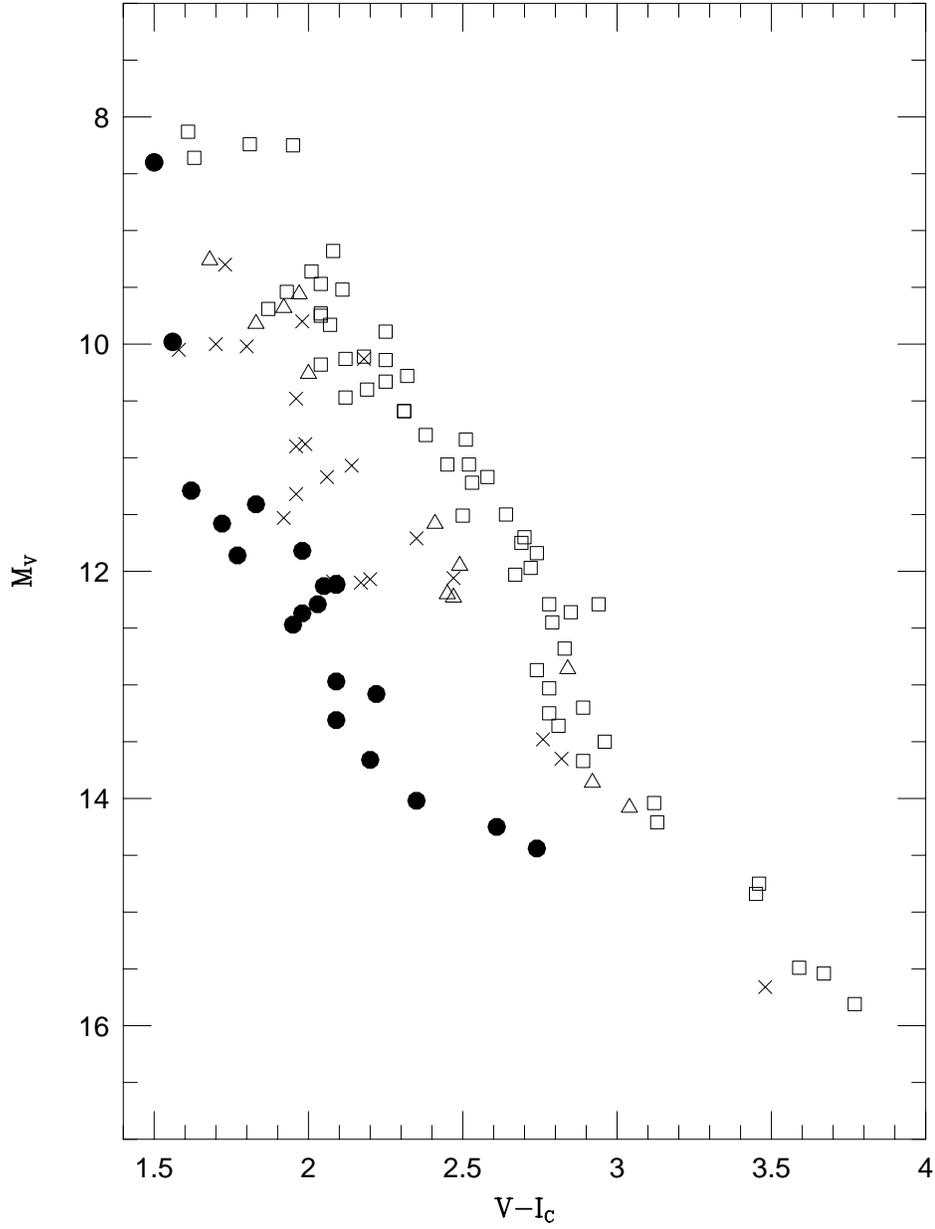}{6in}{0}{70}{70}{-230}{-30}
\caption {The (M$_V$, (V-I)) colour-magnitude diagram for late-type stars. Open
squares identify single stars from the 8-parsec sample (Reid \& Gizis, 1997); 
open triangles are mildly metal-poor disk dwarfs; 
crosses are sdMs; and filled dots are esdM subdwarfs.}
\end{figure}

\begin{figure}
\plotone{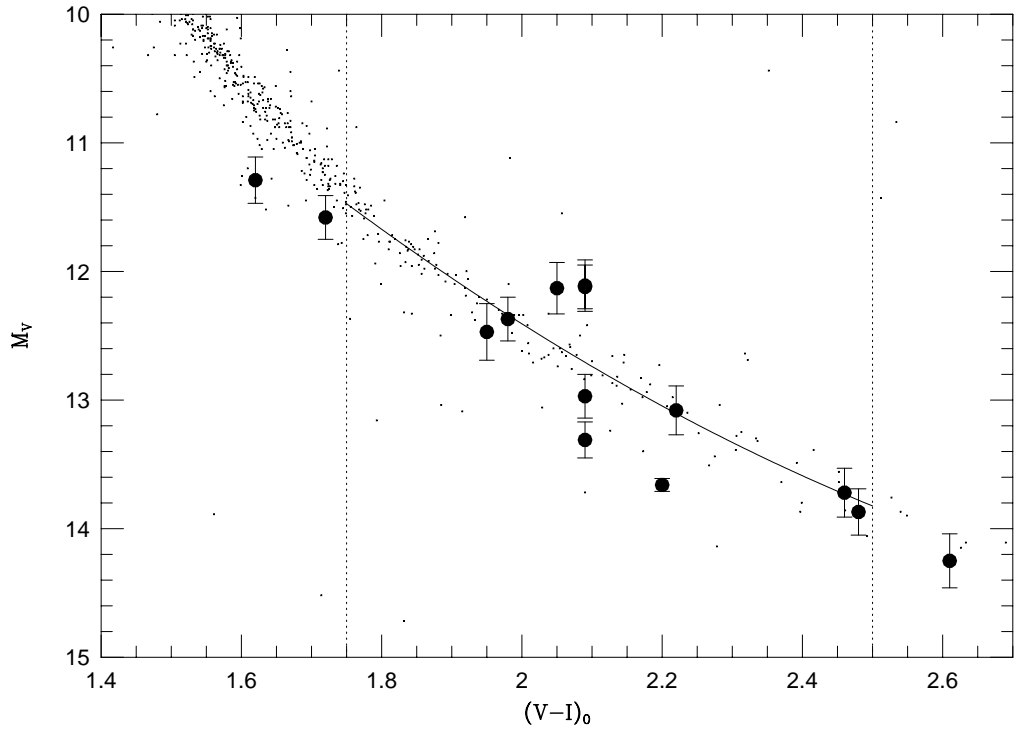}
\caption{ Main-sequence fitting to late-type subdwarfs in NGC 6397. The solid
points are the calibrating esdMs, listed in Table 2; crosses mark data
for proper-motion selected NGC 6397 members. The solid line plots the mean
colour-magnitude relation for the latter stars.}
\end{figure}

\begin{figure}
\plotfiddle{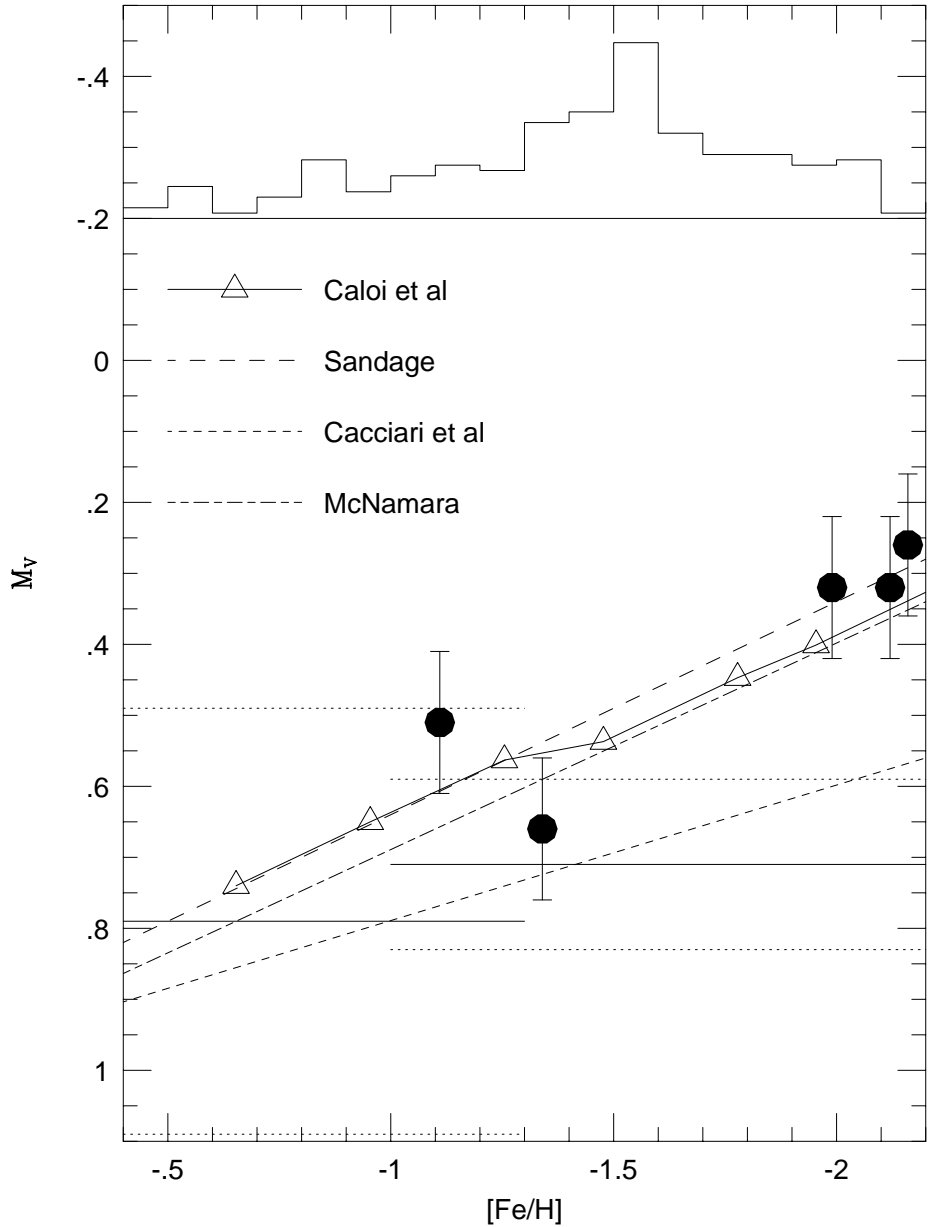}{6in}{0}{70}{70}{-230}{-30}
\caption {The (M$_V$, [Fe/H]) relation for globular cluster RR Lyraes
derived from the data listed in Table 2. The horizontal
solid lines are the statistical parallax solution derived for field stars by 
Layden et al (1996), with the dotted lines marking the $\pm1\sigma$ limits.
Gould \& Popowski's field halo calibration is M$_V=0.77\pm0.15$. 
The linear relations plotted are from Cacciari et al's (1992)
and McNamara's (1997) Baade-Wesselink field-star calibrations,  and from 
Sandage's (1993) cluster variable calibration. The open triangles 
mark the predictions of the Caloi et al (1997) Zero-Age Horizontal 
Branch models. The uppermost histogram plots the abundance distribution 
of the field RR Lyraes analysed by Layden et al.}
\end{figure}


\clearpage
\begin {thebibliography}{}

\bibitem[Adams \& Joy, 1922]{aj22} Adams, W.S., Joy, A.H. 1922, \apj, 56, 242
\bibitem[Alcaino et al, 1997]{al97} Alcaino, G., Liller, W., Alvarado, F., Kravtsov, V.,
Ipatov, A., Samus, N., Smirnov, O. 1997, \aj, 114, 1067
\bibitem[Allard \& Hauschildt, 1995]{ah95} Allard, F., Hauschildt, P.H. 1995, \apj, 445, 433
\bibitem[Anthony-Twarog et al (1992)] {at92} Anthony-Twarog, B.J., Twarog, B.,A., \& Suntzeff, N.B.,
1992, \aj, 103, 1264
\bibitem[Baraffe et al, 1997] {bar97} Baraffe, I., Chabrier, G., Allard, F., Hauschildt, P.H.
1997, \aap, 327, 1054
\bibitem[Barnes \& Hawley, 1987] {bh87} Barnes, T.G., Hawley, S.L. 1987, \apj, 307, L9
\bibitem[Bergbusch \& Vandenberg 1992] {bv92} Bergbusch P.A., Vandenberg, D.A. 1992, \apjs, 81, 163
\bibitem[Bolte \& Hogan (1996)] {bh96} Bolte, M., \& Hogan, C.J. 1995, Nature, 376, 399
\bibitem[Buonnano et al, 1994] {b94} Buonnano, R., Corsi, C.E., Buzzoni, A., Caciari, C., 
Ferraro, F.R., Fusi Pecci, F. 1994, \aap, 290, 69
\bibitem[Cacciari et al, 1992]{cac92} Cacciari, C., Clementini, G., Fernley, J.A. 1992, \apj, 396, 219
\bibitem[Caloi et al]{c97} Caloi, V., D'Antona, F., Mazzitelli, I. 1997, \aap, 320, 823
\bibitem[Cannon, 1974]{c74} Cannon, R.D. 1974, \mnras, 167, 551
\bibitem[Carney et al (1994)] {clla94} Carney, B.W., Latham, D.W., Laird, J.B., \& Aguilar, L.A.
1994, \aj, 107, 2240 
\bibitem[Carretta \& Gratton (1997)] {cg97} Carretta, E. \& Gratton. R.G. 1997, \aaps, 121, 95
\bibitem[Cassisi et al, 1998] {cas98} Cassisi, S., Castellani, V., Degl'Innocenti, S., 
Weiss, A. 1998, \aaps, 129, 267
\bibitem[Chaboyer et al (1998)] {cds96} Chaboyer, B., Demarque, P., Kernan, P.J., Krauss, L.M. 
1998, \apj, 494, 96
\bibitem[Cool et al, 1996]{cpk96} Cool, A.M., Piotto, G., King, I.R. 1996, \apj, 468, 655
\bibitem[Cottrell, 1978]{c78} Cottrell, P.L. 1978, \apj, 223, 544
\bibitem[Dahn et al, 1995]{dah95} Dahn, C.C., Liebert, J., Harris, H.C., Guetter, H.H. 1995, in The Bottom
of the Main Sequence - and beyond, ed. by C.G. Tinney (Berlin, Springer), p. 239
\bibitem[D'Antona et al (1997)] {da97} D'Antona, F., Caloi, V., \& Mazzitelli, I. 1997, \apj, 477, 519 
\bibitem[Drukier et al, 1993]{dk93} Drukier, G.A., Fahlman, G.G., Richer, H.B., Searle, L., Thompson, I.
1993, \aj, 106, 2335
\bibitem[ESA, 1997] {esa} ESA, 1997, The Hipparcos catalogue, ESA SP-1200
\bibitem[Fahlman et al, 1989]{fah89} Fahlman, G.G., Richer, H.B., Searle, L., Thompson, I.B. 1989,
\apj, 343, L49
\bibitem[Feast, 1997] {f97} Feast, M.W. 1997, \mnras, 284, 761
\bibitem[Fernley et al (1998)] {fer98a} Fernley, J.A., Barnes, T.G., Skillen, I., Hawley, S.L., Hanley, C.J.,
Evans, D.W., Solano, E., Garrido, R. 1998, \aap, 330, 515
\bibitem[Gizis, 1997]{g97} Gizis, J.E. 1997, \aj, 113, 806
\bibitem[Gizis \& Reid, 1997]{g97a} Gizis, J.E., Reid, I.N. 1997a, \pasp, 109, 849
\bibitem[Gizis \& Reid, 1997]{g97b} Gizis, J.E., Reid, I.N. 1997b, \pasp, 109, 1233
\bibitem[Gould \& Popowski, 1998] {gp98} Gould, A., Popowski, P. 1998, \apj, in press
\bibitem[Graham \& Doremus]{gd68} Graham, J.A., Doremus, C. 1968, \aj, 73, 226
\bibitem[Gratton et al (1997b)] {gf97} Gratton, R.G., Fusi Pecci, F., Carretta, E., 
Clementini, G., Corsi, C.E., \& Lattanzi, M. 1997, \apj, 491, 749
\bibitem[Greenstein, 1971]{g71} Greenstein, J.E. 1971, in IAU Symposium 41, {\sl White Dwarfs}, ed. W.J. Luyten,
(Dordrecht:Reidel) p.46
\bibitem[He et al, 1995]{he95} He, L.D., Whittet, D.C.B., Kilkenny, D., Jones, J.H.S 1995, \apjs, 101, 335
\bibitem[Heber \& Kudritzki, 1986]{hk86} Heber, U., Kudritzki, R.P. 1986, \aap, 169, 244
\bibitem[Holtzman et al, 1995]{h95} Holtzman, J.A., Burrows, C.J., Casertano, C., Hester, J.J.,
Trauger, J.T., Watson, A.M., Worthey, G. 1995, \pasp, 107, 1065
\bibitem[Jones, 1973]{j73} Jones, D.H.P. 1973, \mnras, 161, 19P
\bibitem[King et al, 1998] {ki98} King, I.R., Anderson, J., Cool, A.M., Piotto, G. 1998, \apj, 492, L37
\bibitem[King et al, 1998] {ki98a}King, J.R., Stephens, A., Boesgaard, A.M., Deliyaniis, C.P. 1998,
\aj, 115, 666
\bibitem[Kuiper, 1939]{k39} Kuiper, G.P. 1939, \apj, 89, 549
\bibitem[Laird et al (1988)] {lcl88} Laird, J.B., Carney, B.W., \& Latham, D.W. 1988, \aj, 95, 1843
\bibitem[Layden et al (1996)] {l96} Layden, A.C., Hanson, R.B., Hawley, S.L., Klemola, A.R.,
\& Hanley, C.J. 1996, \aj, 112, 2110
\bibitem[Lee et al, 1990] {le90} Lee, Y.W., Demarque, P., Zinn, R. 1990, \apj, 350, 155
\bibitem[Lutz \& Kelker (1973)] {lk73} Lutz, T.E., \& Kelker, D.H. 1973, \pasp, 85, 573
\bibitem[Luyten, 1980]{l80} Luyten, W.J. 1980, The LHS Catalogue, Univ. of Minnesota, Minneapolis
\bibitem[McNamara, 1992]{mc97} McNamara, D.H. 1997, \pasp, 109, 857
\bibitem[Monet et al, 1992] {mon92} Monet, D.G., Dahn, C.C., Vrba, F.J., Harris, H.C., Pier, J.R., 
Luginbuhl, C.B., Ables, H.D. 1992, \aj, 103, 638
\bibitem[Mould, 1976]{m76} Mould, J.R. 1976, \apj, 207, 535
\bibitem[Newell, et al, 1969] {n69} Newell, E.B., Rodgers, A.W., Searle, L. 1969, \apj, 156, 597
\bibitem[Paresce et al, 1995]{pd95} Paresce, F., Demarchi, G., Romaniello, M. 1995, \apj, 440, 216
\bibitem[Pont et al (1997)] {po97} Pont, F., Mayor, M., Turon, C., Vandenberg, D.A. 1998,
\aap, 329, 87
\bibitem[Press et al]{pres} Press, W.H., Teukolsky, S.A., Vetterling, W.T., \& Flannery, B.P. 1992,
Numerical Recipes, 2nd Ed.  Cambridge University Press.
\bibitem[Reid (1996)] {re96} Reid, I.N. 1996, \mnras, 278, 367
\bibitem[Reid (1997)] {re97} Reid, I.N. 1997, \aj, 114, 161 
\bibitem[Reid (1998)] {re98} Reid, I.N. 1998, \aj, 115, 161 
\bibitem[Reid \& Gizis, 1997]{rg97} Reid, I.N., Gizis, J.E. 1997, \aj, 113, 2246
\bibitem[Reid et al, 1995] {red95}  Reid, I.N., Hawley, S.L., Gizis, J.E. 1995 \aj, 110, 1838
\bibitem[Ruiz \& Anguita, 1993] {ra93} Ruiz, M.T., Anguita, C. 1993, \aj, 105, 614
\bibitem[Sandage \& Eggen 1959] {se59} Sandage, A., Eggen, O.J. 1959, \mnras, 119, 278
\bibitem[Sandage 1970] {sa70} Sandage, A. 1970, \apj, 162, 841
\bibitem[Sandage 1982] {sa82} Sandage, A. 1982, \apj, 252, 533
\bibitem[Sandage 1993] {sa93} Sandage, A. 1993, \aj, 106, 703
\bibitem[Schlegel et al, 1998]{sc98} Schlegel, D.J., Finkbeiner, D.P., \& Davis, M.  1998, \apj, 500, 525
\bibitem[Silbermann \& Smith]{ss95} Silbermann, N., Smith, H.A. 1995, \aj, 110, 704
\bibitem[Strugnell et al, 1987] {st87} Strugnell, P., Reid, I.N., Murray, C.A. 1986, \mnras, 220, 413
\bibitem[Udalski, 1998]{u98} Udalski, A. 1998, Acta. Astr. in press
\bibitem[Van Altena et al, 1991] {va91} Van Altena, W.F., Lee, J.T. \& Hoffleit,
E.D. 1996, The General Catalog of Trigonometric  Parallaxes, Yale Univ.
\bibitem[van den Bergh (1988)]{vdb88} van den Bergh, S. 1988, \aj, 95, 106
\bibitem[Vandenberg et al (1990)] {vbs90} Vandenberg, D.A., Bolte, M., \& Stetson, P.B. 1990, \aj, 
100, 445 (VBS)
\bibitem[Walker, 1994] {w94} Walker, A. 1994, \aj, 108, 555

\end{thebibliography}
\end{document}